\documentclass[Physsubmission, Phys]{SciPost}

\hypersetup{
    colorlinks,
    linkcolor={red!50!black},
    citecolor={blue!50!black},
    urlcolor={blue!80!black}
}

\usepackage[bitstream-charter]{mathdesign}
\DeclareSymbolFont{usualmathcal}{OMS}{cmsy}{m}{n}
\DeclareSymbolFontAlphabet{\mathcal}{usualmathcal}

\begin{document}

\begin{center}{\Large \textbf{
Phenomenological study on correlation between flow harmonics and mean transverse momentum in nuclear collisions\\
}}\end{center}

\begin{center}
Chunjian Zhang\textsuperscript{1$\star$},
Jiangyong Jia\textsuperscript{1,2} and
Shengli Huang\textsuperscript{1}
\end{center}

\begin{center}
{\bf 1} Department of Chemistry, Stony Brook University, Stony Brook, NY 11794, USA
\\
{\bf 2} Physics Department, Brookhaven National Laboratory, Upton, NY 11976, USA
\\
* chun-jian.zhang@stonybrook.edu
\end{center}

\begin{center}
\today
\end{center}


\definecolor{palegray}{gray}{0.95}
\begin{center}
\colorbox{palegray}{
  \begin{tabular}{rr}
  \begin{minipage}{0.1\textwidth}
    \includegraphics[width=30mm]{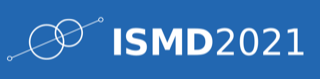}
  \end{minipage}
  &
  \begin{minipage}{0.75\textwidth}
    \begin{center}
    {\it50th International Symposium on Multiparticle Dynamics}\\{\it(ISMD2021)}\\
    {\it12-16 July 2021} \\
    \doi{10.21468/SciPostPhysProc.?}\\
    \end{center}
  \end{minipage}
\end{tabular}
}
\end{center}

\section*{Abstract}
{\bf
To assess the properties of the quark-gluon plasma formed in nuclear collisions, the Pearson correlation coefficient between flow harmonics and mean transverse momentum, $\rho\left(v_{n}^{2},\left[p_{\mathrm{T}}\right]\right)$, reflecting the overlapped geometry of colliding atomic nuclei, is measured. $\rho\left(v_{2}^{2},\left[p_{\mathrm{T}}\right]\right)$ was found to be particularly sensitive to the quadrupole deformation of the nuclei. We study the influence of the nuclear quadrupole deformation on $\rho\left(v_{n}^{2},\left[p_{\mathrm{T}}\right]\right)$  in $\rm{Au+Au}$ and $\rm{U+U}$ collisions at RHIC energy using $\rm{AMPT}$ transport model, and show that the $\rho\left(v_{2}^{2},\left[p_{\mathrm{T}}\right]\right)$ is reduced by the quadrupole deformation $\beta_2$ and turns to change sign in ultra-central collisions (UCC). 
}

\section{Introduction}
\label{sec:intro}
A hot and dense phase of Quantum Chromodynamics (QCD) matter, so-called quark-gluon plasma (QGP) \cite{doi:10.1146/annurev-nucl-101918-023825}, is naturally created in the nuclear collisions studied at the Relativistic Heavy Ion Collider (RHIC) and the Large Hadron Collider (LHC). The QGP expands due to pressure gradients between the medium and the outside vacuum. During this expansion, the initial spatial anisotropies lead to final-state momentum anisotropies. The large azimuthal modulations in the final distributions of the produced particles are typically characterized as a Fourier series \cite{Voloshin:1994mz}: $\frac{d N}{d \phi} \propto 1+2 \Sigma_{n=1}^{\infty} v_{n} \cos \left(n\left(\phi-\Phi_{n}\right)\right)$, where $v_n$ and $\Phi_n$ represent the magnitude and event-plane angle of the $n^\mathrm{th}$-order harmonic flow. Interestingly, the shape and orientation of the deformed nuclei characterized by the nuclear deformation, if taken into account for the generation of the initial state geometry, could result in the non-trivial behavior of final-state bulk observables \cite{PhysRevLett.127.242301,PhysRevLett.128.022301,Jia:2021oyt}.

Recently, the correlation between flow harmonics ($v_{n}$) and mean transverse momentum ($[p_T]$), $\rho(v_{n}\{2\}^2,[p_T])$ \cite{Bozek:2020drh}, was proposed to be sensitive to distinguish the nuclear deformation \cite{Giacalone:2020awm}. Of particular interest is a significantly negative correlation in central U+U collisions at STAR Collaboration \cite{isJJia} due to the quadrupole deformation $\beta_2$ where the $\beta_2$ values are obtained from the measured reduced electric transition probability $B(E_n)\uparrow$ via the standard formula $\beta_{2}=\frac{4 \pi}{3 Z R_{0}^{2}} \sqrt{\frac{B(E 2) \uparrow}{e^{2}}}$ \cite{bohr}. Comparison with the $\beta_2$ scan in phenomenological study can explore the sensitivity of $\rho\left(v_{n}^{2},\left[p_{\mathrm{T}}\right]\right)$ to the fluctuations in the initial geometry arising from nuclei shape at a much shorter time scale ($\sim 10^{-24}$) in relativistic heavy-ion collisions. To further constrain the initial conditions and transport properties in hydrodynamic evolution, ATLAS \cite{ATLAS:2019pvn,isAra} and ALICE Collaboration \cite{ALICE:2021gxt} have also reported this measurements in system scan $pp$, $p$+Pb, Xe+Xe, and Pb+Pb collisions. 

In this proceeding, The influences of the nuclear quadrupole deformation on $\rho\left(v_{n}^{2},\left[p_{\mathrm{T}}\right]\right)$ in Au+Au and U+U collisions at $\sqrt{s_{NN}}=200$ GeV are studied with the framework of AMPT transport model.

\section{Analysis}
The $\rho\left(v_{n}^{2},\left[p_{\mathrm{T}}\right]\right)$ is quantified by a three-particle correlator defined as:

\begin{equation}
\rho\left(v_{n}^{2},\left[p_{\mathrm{T}}\right]\right)=\frac{\left\langle v_{n}^{2} \delta p_{\mathrm{T}}\right\rangle}{\sqrt{\left\langle\left(\delta v_{n}^{2}\right)^{2}\right\rangle\left\langle\delta p_{\mathrm{T}} \delta p_{\mathrm{T}}\right\rangle}} = \frac{\left\langle\frac{\sum_{i, j, k, i \neq j \neq k} e^{i n\left(\phi_{i}-\phi_{j}\right)}\left(p_{\mathrm{T}, k}-\left\langle\left[p_{\mathrm{T}}\right]\right\rangle\right)}{\sum_{i, j, k, i \neq j \neq k}}\right\rangle}{\sqrt{\left(\left\langle v_{n}^{4}\right\rangle-\left\langle v_{n}^{2}\right\rangle^{2}\right) \left\langle\frac{\sum_{i, j, i \neq j}\left(p_{\mathrm{T}, i}-\left\langle\left[p_{\mathrm{T}}\right]\right\rangle\right)\left(p_{\mathrm{T}, j}-\left\langle\left[p_{\mathrm{T}}\right]\right\rangle\right)}{\sum_{i, j, i \neq j}}\right)}}
\end{equation}
where the indices $i$, $j$ and $k$ loop over distinct particles to account for all unique triplets, and the $\left\langle \right\rangle$ denotes average over events. In this analysis, we use all particles within $|\eta|<$ 2 and 0.2 $< p_T <$ 2 GeV/c for the benefit of statistical precision and the centrality is defined using the number of participants $\rm{N_{part}}$. We use the AMPT model v2.26t5 \cite{PhysRevC.72.064901} with string-melting mode and partonic cross section of 3.0 mb, which we check reasonably reproduce Au+Au flow data at RHIC. The systematic study of short-range $``$non-flow$"$ effect from resonance-decays, jets and dijets was checked in Ref. \cite{ZHANG2021136702}.

\section{Result}
\label{sec:results}

\begin{figure}[h]
\centering
\includegraphics[width=0.75\textwidth]{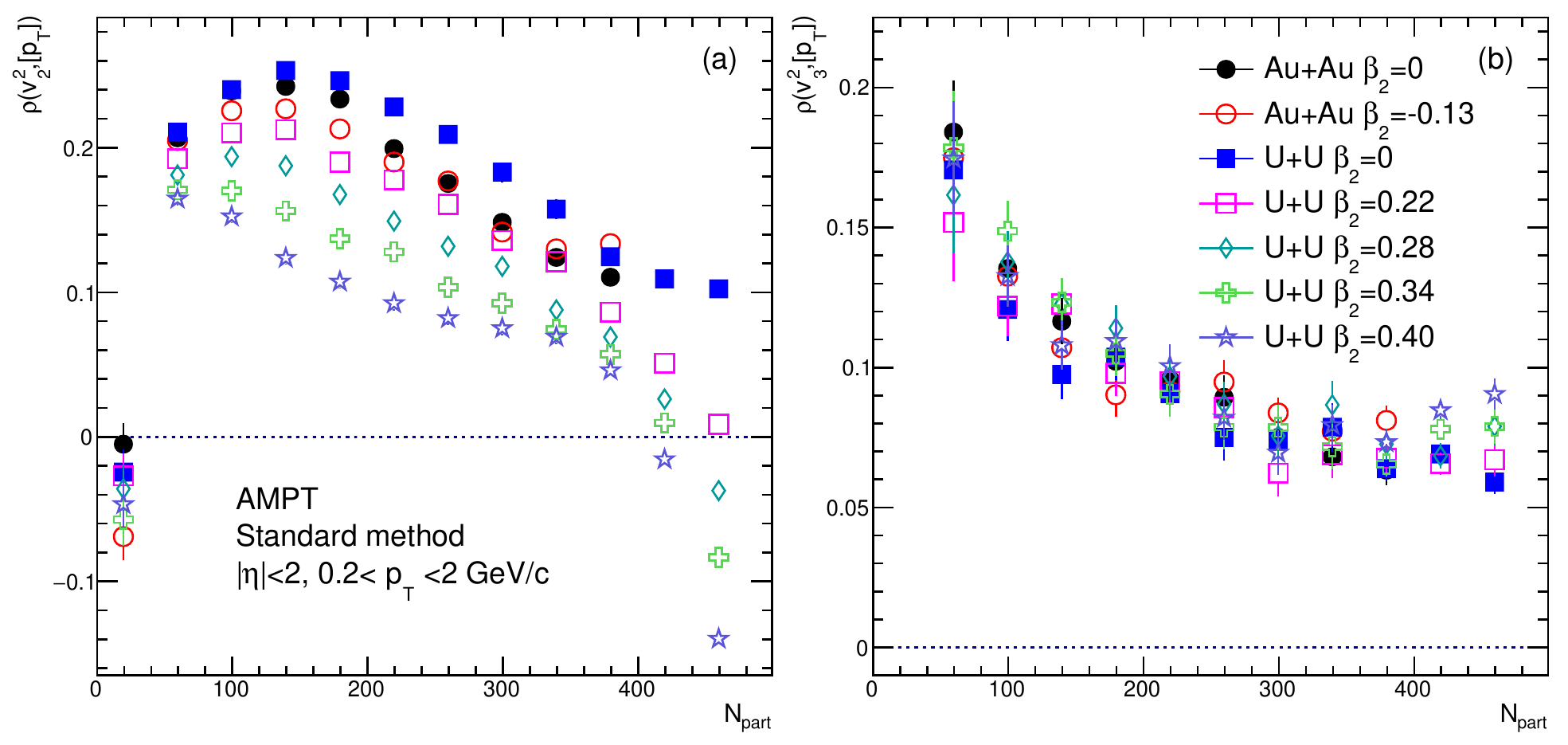}
\caption{The $\rm{N_{part}}$ dependence of the Pearson correlation coefficient $\rho\left(v_{n}^{2},\left[p_{\mathrm{T}}\right]\right)$ for n = 2 (left) and 3 (right) in Au+Au and U+U collisions with different deformation parameter $\beta_2$.}
\label{fig:1}
\end{figure}

Figures \ref{fig:1} shows the Pearson correlation coefficients $\rho\left(v_{n}^{2},\left[p_{\mathrm{T}}\right]\right)$ for n=2 and 3 calculated in standard method using final-state hadrons in Au+Au and U+U collisions with different $\beta_2$. The $\rho\left(v_{2}^{2},\left[p_{\mathrm{T}}\right]\right)$ shows strong non-trivial dependence on $\beta_2$.  In particular, we observe the strongest sensitivity in the UCC region, where a large positive $\beta_2$ leads to a negative $\rho\left(v_{2}^{2},\left[p_{\mathrm{T}}\right]\right)$ due to the orientation of the colliding deformed nuclei. In the mid-central and peripheral regions, the magnitudes of $\rho\left(v_{2}^{2},\left[p_{\mathrm{T}}\right]\right)$ always decrease with increasing magnitude of $\beta_2$. The $\rho\left(v_{3}^{2},\left[p_{\mathrm{T}}\right]\right)$ are positive for both collision systems since triangular flow $v_3$ is purely fluctuation driven which is insensitive to the nuclear geometric effect. The detailed study can be found in Ref. \cite{PhysRevC.105.014906}.

\section{Conclusion}
We studied the influence of the nuclear quadrupole deformation on the $\rho\left(v_{n}^{2},\left[p_{\mathrm{T}}\right]\right)$ in Au+Au and U+U collisions at RHIC energy using the AMPT transport model. $\rho\left(v_{2}^{2},\left[p_{\mathrm{T}}\right]\right)$ shows a strong dependence on $\beta_2$ and turns to change sign in the ultra-central collisions, while the $\rho\left(v_{3}^{2},\left[p_{\mathrm{T}}\right]\right)$ are always similar and positive. Detailed comparison of the model prediction with the results from STAR experimental data in Au+Au and U+U collisions could allow us to constrain the $\beta_2$ values of uranium nucleus.

\section*{Acknowledgements}
This work is supported by DOE DEFG0287ER40331 and NSF PHY-1913138.
\bibliography{ISMDProceeding.bib}

\begin{thebibliography}{10}
\providecommand{\url}[1]{\texttt{#1}}
\providecommand{\urlprefix}{URL }
\expandafter\ifx\csname urlstyle\endcsname\relax
  \providecommand{\doi}[1]{doi:\discretionary{}{}{}#1}\else
  \providecommand{\doi}{doi:\discretionary{}{}{}\begingroup
  \urlstyle{rm}\Url}\fi
\providecommand{\eprint}[2][]{\url{#2}}

\bibitem{doi:10.1146/annurev-nucl-101918-023825}
S.~Schlichting and D.~Teaney,
\newblock \emph{The first fm/c of heavy-ion collisions},
\newblock Annual Review of Nuclear and Particle Science \textbf{69}(1), 447
  (2019),
\newblock \doi{10.1146/annurev-nucl-101918-023825},
\newblock \eprint{https://doi.org/10.1146/annurev-nucl-101918-023825}.

\bibitem{Voloshin:1994mz}
S.~Voloshin and Y.~Zhang,
\newblock \emph{{Flow study in relativistic nuclear collisions by Fourier
  expansion of Azimuthal particle distributions}},
\newblock Z. Phys. C \textbf{70}, 665 (1996),
\newblock \doi{10.1007/s002880050141},
\newblock \eprint{hep-ph/9407282}.

\bibitem{PhysRevLett.127.242301}
G.~Giacalone, J.~Jia and C.~Zhang,
\newblock \emph{Impact of nuclear deformation on relativistic heavy-ion
  collisions: Assessing consistency in nuclear physics across energy scales},
\newblock Phys. Rev. Lett. \textbf{127}, 242301 (2021),
\newblock \doi{10.1103/PhysRevLett.127.242301}.

\bibitem{PhysRevLett.128.022301}
C.~Zhang and J.~Jia,
\newblock \emph{Evidence of quadrupole and octupole deformations in
  $^{96}\mathrm{Zr}+^{96}\mathrm{Zr}$ and $^{96}\mathrm{Ru}+^{96}\mathrm{Ru}$
  collisions at ultrarelativistic energies},
\newblock Phys. Rev. Lett. \textbf{128}, 022301 (2022),
\newblock \doi{10.1103/PhysRevLett.128.022301}.

\bibitem{Jia:2021oyt}
J.~Jia and C.-J. Zhang,
\newblock \emph{{Scaling approach to nuclear structure in high-energy heavy-ion
  collisions}}  (2021),
\newblock \eprint{2111.15559}.

\bibitem{Bozek:2020drh}
P.~Bozek and H.~Mehrabpour,
\newblock \emph{{Correlation coefficient between harmonic flow and transverse
  momentum in heavy-ion collisions}},
\newblock Phys. Rev. C \textbf{101}(6), 064902 (2020),
\newblock \doi{10.1103/PhysRevC.101.064902},
\newblock \eprint{2002.08832}.

\bibitem{Giacalone:2020awm}
G.~Giacalone,
\newblock \emph{{Constraining the quadrupole deformation of atomic nuclei with
  relativistic nuclear collisions}},
\newblock Phys. Rev. C \textbf{102}(2), 024901 (2020),
\newblock \doi{10.1103/PhysRevC.102.024901},
\newblock \eprint{2004.14463}.

\bibitem{isJJia}
J.~Jia,
\newblock
  \emph{\href{https://indico.cern.ch/event/854124/contributions/4135480/attachments/2170549/3665249/JiangyongJia_01_14_2021_v5.pdf}{Nuclear
  deformation effects via Au+Au and U+U collisions from STAR}} (Initial Stage
  2021).

\bibitem{bohr}
A.~Bohr and B.~R. Mottelson, eds.,
\newblock \emph{{Nuclear Structure}},
\newblock World Scientific,
\newblock ISBN 978-981-238-660-1,
\newblock \doi{https://doi.org/10.1142/3530} (1998).

\bibitem{ATLAS:2019pvn}
G.~Aad \emph{et~al.},
\newblock \emph{{Measurement of flow harmonics correlations with mean
  transverse momentum in lead-lead and proton-lead collisions at
  $\sqrt{s_{NN}}=5.02$ TeV with the ATLAS detector}},
\newblock Eur. Phys. J. C \textbf{79}(12), 985 (2019),
\newblock \doi{10.1140/epjc/s10052-019-7489-6},
\newblock \eprint{1907.05176}.

\bibitem{isAra}
A.~Behera,
\newblock
  \emph{\href{https://indico.cern.ch/event/854124/contributions/4134918/attachments/2169460/3663379/IS21_Arabinda_v3.pdf}{Correlations
  between flow and transverse momentum in Pb+Pb and Xe+Xe collisions with
  ATLAS}} (Initial Stage 2021).

\bibitem{ALICE:2021gxt}
S.~Acharya \emph{et~al.},
\newblock \emph{{Characterizing the initial conditions of heavy-ion collisions
  at the LHC with mean transverse momentum and anisotropic flow correlations}}
  (2021),
\newblock \eprint{2111.06106}.

\bibitem{PhysRevC.72.064901}
Z.-W. Lin, C.~M. Ko, B.-A. Li, B.~Zhang and S.~Pal,
\newblock \emph{Multiphase transport model for relativistic heavy ion
  collisions},
\newblock Phys. Rev. C \textbf{72}, 064901 (2005),
\newblock \doi{10.1103/PhysRevC.72.064901}.

\bibitem{ZHANG2021136702}
C.~Zhang, A.~Behera, S.~Bhatta and J.~Jia,
\newblock \emph{Non-flow effects in correlation between harmonic flow and
  transverse momentum in nuclear collisions},
\newblock Physics Letters B \textbf{822}, 136702 (2021),
\newblock \doi{https://doi.org/10.1016/j.physletb.2021.136702}.

\bibitem{PhysRevC.105.014906}
J.~Jia, S.~Huang and C.~Zhang,
\newblock \emph{Probing nuclear quadrupole deformation from correlation of
  elliptic flow and transverse momentum in heavy ion collisions},
\newblock Phys. Rev. C \textbf{105}, 014906 (2022),
\newblock \doi{10.1103/PhysRevC.105.014906}.

\end{thebibliography}

\nolinenumbers

\end{document}